\documentclass[draft]{agujournal2018}
\usepackage{apacite}
\usepackage{url} 
\usepackage{lineno}
\usepackage{ragged2e}
\usepackage[pass,letterpaper]{geometry}
\journalname{Geophysical Research Letters}

\begin{document}


\justifying

\title{Visibility and Line-Of-Sight Extinction Estimates in Gale Crater during the 2018/MY34 Global Dust Storm}

\authors{Christina L. Smith\affil{1}, John E. Moores\affil{1}, Mark Lemmon\affil{2}, Scott D. Guzewich\affil{3}, Casey A. Moore\affil{1}, Douglas Ellison\affil{4}, Alain S. J. Khayat\affil{3,5}}

\affiliation{1}{York University, 4700 Keele Street, Toronto, ON, M3J 1PL, Canada.}
\affiliation{2}{Space Science Institute - 4750 Walnut St, Suite 205, Boulder, CO 80301}
\affiliation{3}{NASA Goddard Space Flight Center, 8800 Greenbelt Rd, Greenbelt, MD, 20771, USA.}
\affiliation{4}{NASA Jet Propulsion Laboratory, 4800 Oak Grove Drive, Pasadena, CA, 91109, USA.}
\affiliation{5}{Center for Research and Exploration in Space Science \& Technology (CRESST II), Department of Astronomy, University of Maryland, College Park, MD 20742, USA.}

\correspondingauthor{Christina L. Smith}{chrsmith@yorku.ca}


\begin{keypoints}
\item Line-of-sight extinction during the global dust storm peaks an order of magnitude higher than previously observed maxima at Gale Crater.
\item Northern and western directions show a peak extinction then initial decrease, followed by a secondary peak prior to the final decay.
\item Morning extinction values are systematically higher than afternoon values, potentially indicative of atmospheric mixing. 
\end{keypoints}

\begin{abstract}

Northern line-of-sight extinction within Gale Crater during the 2018 global dust storm was monitored daily using MSL's Navcam. Additional observations with Mastcam (north) and Navcam (all directions) were obtained at a lower cadence. Using feature identification and geo-referencing, extinction was estimated in all possible directions. Peak extinction of $>1.1$ km$^{-1}$ was measured between sols 2086 and 2090, an order of magnitude higher than previous maxima. Northern and western directions show an initial decrease, followed by a secondary peak in extinction, not seen in column opacity measurements. Due to foreground topography, eastern direction results are provided only as limits, and southern results were indeterminable. Mastcam red and green filter results agree well, but blue filter results show higher extinctions, likely due to low signal-to-noise. Morning results are systematically higher than afternoon results, potentially indicative of atmospheric mixing. 

\end{abstract}


\section*{Plain Language Summary}

Once roughly every three Mars Years (5.5 Earth Years), Mars undergoes a dust storm so large it wraps around the planet, called a global dust storm. One of these storms occurred in 2018. The Curiosity Rover was able to operate throughout the storm and provided a unique view from the surface. During the 2018 storm, Curiosity obtained daily images which were used to measure how far the rover could see in one or more directions and, from that, monitor the dust within Gale Crater. At the peak of the storm, visibility was reduced to less than 3 km - for comparison the crater rim, which sits approximately 30 km away, is visible at most times of year. In northern and western directions, visibility improved temporarily 15-20  Mars Days after the peak of the storm before worsening again and then slowly improving through to the end of the storm. Foreground features blocked sections in the east and south from view so only constraints could be placed on the eastern directions and nothing could be determined for the south. Morning visibility was worse than afternoon/evening visibility, which might be caused by atmospheric mixing processes.


\section{Introduction}

The landing site of the Mars Science Laboratory (MSL) Rover is Gale Crater, a 154 km wide impact crater with 5 km high \emph{Aeolis Mons} in the center. Atmospheric mixing is strongly affected by the topography, observable in the aerosol abundance distribution within and above Gale. MSL monitors the line-of-sight (LOS) extinction within \citep[e.g.][]{Moore2016} and column opacity above \citep[e.g.][]{Smith2016} Gale Crater to determine variations in aerosol abundance and dust settling. The former has been consistently monitored by MSL since sol 100 \citep{Moores2015, Moore2016, Moore2019} and has shown a strong, repeating seasonal pattern with peak amplitude of 0.06 km$^{-1}$, and average peak values of 0.11 km$^{-1}$. 

Martian local/regional dust storms are common during the southern hemisphere summer seasons. However, global dust storms (GDSs), where regional storms grow or coalesce to become planet-encircling, are rare phenomena and occur on average once every three Mars Years \citep{Zurek1993}. GDSs loft dust $>60$ km into the atmosphere \citep[e.g.][and references therein]{Clancy2010} and take several months to dissipate. GDSs have significant effects on atmospheric features such as diurnal pressure variations, circulation, and atmospheric heating \citep[e.g.][]{Guzewich2018, Viudez2018}. However, the series of events that lead to the formation of a GDS are not well known. Given the rarity and unpredictability of these events coupled with our limited understanding of their formation, it is vital that any opportunity to study GDSs be exploited.

As detailed in \citet{Guzewich2018}, prior to the onset of the 2018/MY34 GDS, the MSL team prepared a cadence of atmospheric observations that would be initiated if potential GDS-precursor conditions were detected (e.g. atmospheric warming, rapid opacity increase). This campaign included a daily north-facing horizon image (LOS image), taken around local noon with MSL's Engineering and Navigation cameras (Navcam). This style of image has been regularly used to monitor the LOS extinction between MSL and the crater rim, approximately 30 km away \citep[e.g.][]{Moores2015, Moore2016}. The campaign also included an increased cadence of Mastcam Crater Rim Extinction (CRE) observations which also point towards the northern horizon but image in three color filters and have a smaller field of view (FOV). On sol 2075, the GDS campaign was initiated and persisted for 100 sols. 

In this paper, the maximum visible distance and approximate LOS extinction throughout the 2018/MY34 GDS was determined both as a function of time and direction to better understand the evolution of the GDS and its effect on dust lifting and settling within Gale Crater. This paper provides the first estimates of LOS extinction taken from the surface of Mars throughout a GDS. 

\section{Observations}

\begin{figure}
\includegraphics[width=\textwidth]{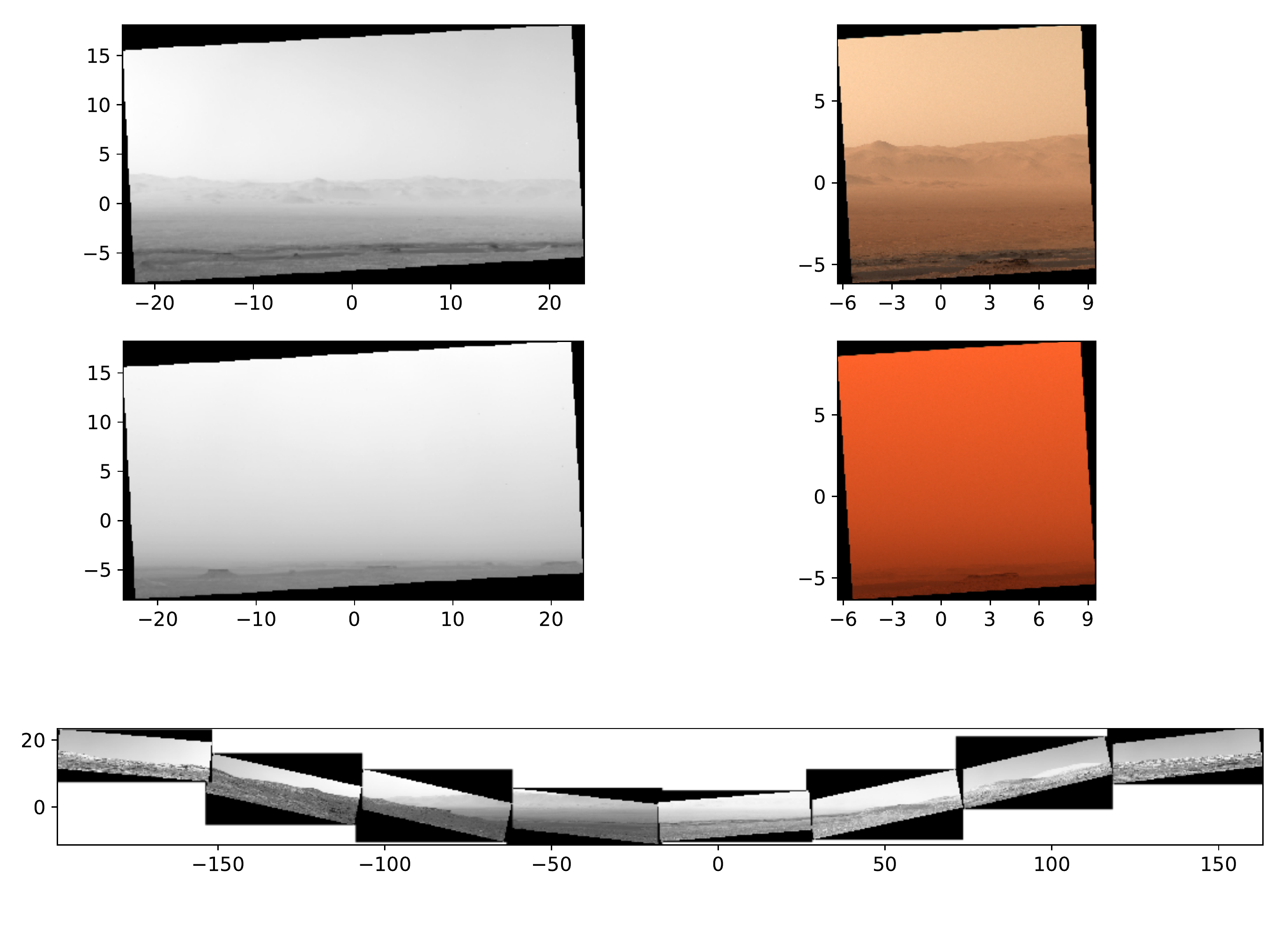}
\caption{Examples of all observations used in this study. The left and right images are Navcam LOS and Mastcam CRE images respectively, taken on sol 2075 (upper) and 2085 near the visibility minimum (center). The lower image is a Navcam DD360 mosaic taken on sol 2074. Axes on all figures show azimuth (abscissa) and elevation (ordinate) in degrees.}\label{exampleobs} 
\end{figure}

All images used in this paper are taken with one of two camera systems: Navcam and Mastcam. Navcam, mounted on MSL's mast has an approximate $45^\circ\times45^\circ$ FOV, and provides a radiometrically calibrated image with a central wavelength of $\sim650$ nm. All Navcam images used within this paper are the most recent Reduced Data Records (RDRs) available. MSL's Mastcam-34 has a maximum $20^\circ\times 15^\circ$ FOV and was used with the clear filter. Examples of each observation are shown in Figure \ref{exampleobs}. 

\subsection{Navcam LOS images}

Navcam LOS images use a $1024\times 512$ (width $\times$ height) pixel sub-frame downsampled onboard to $512\times 256$ pixels, the equivalent of a $45^\circ \times 23^\circ$ degree FOV image. During the GDS, Navcam LOS images have approximate Az/El pointings of 0$^\circ$/5$^\circ$ and contain regions of the sky, crater rim, and ground. Image rotation is corrected using the metadata-supplied quaternions. 97 LOS images were obtained during the GDS campaign. 

\subsection{Navcam DD360 frames}

Navcam dust devil 360 movies (DD360), usually used to monitor dust devil activity \citep[e.g.][] {Lemmon2017}, are sets of three images taken at eight different azimuthal pointings (total of 24 frames). Each frame of the DD360 movie is a $1024\times 256$  pixel sub-frame downsampled onboard to $512\times 128$, the equivalent of a $45^\circ \times 11^\circ$ degree FOV image.  By taking one frame from each azimuthal direction, a $360^\circ$ view of the horizon can be constructed. These are corrected for rotation as with the Navcam LOS images. 242 DD360 frames were used from the 30 DD360 observations obtained during the GDS. 

\subsection{Mastcam CRE images}

Mastcam CRE images use a $1184\times 1184$ pixel sub-frame (the equivalent of a $11^\circ \times 11^\circ$ FOV) with approximate pointings of $1.5^\circ$/$1.7^\circ$ Az/El. 66 Mastcam CREs were obtained during the GDS. Unlike the Navcam images, the Mastcam archived RDRs are intended for quick-look purposes only \citep{Bell2017}. Therefore for this work we use the most recent losslessly downlinked Experimental Data Records (EDRs) and follow the calibration method of \citet{Bell2017}, \citet{Lemmon2015}, and summarized below.

EDRs were decompanded from 8 bits per pixel to 11 bits per pixel using Table App.B of \citet{Bell2017}. 3 Data Numbers (DN), equivalent to the dark current, were subtracted from each image prior to division by updated sky-flats (sol 320), clipped to the image dimensions. The flat-field-corrected image was then clipped to $1184\times 1184$ pixels to remove areas affected by vignetting and debayered  using a \citet{Malvar2004} algorithm, resulting in three separate frames for the red, green, and blue filters (effective wavelengths: 640, 554 and 495 nm respectively). 

Hot and gray pixels as per Table 6 of \citet{Bell2017}, with the eight pixels surrounding each, were removed from each frame and replaced with the mean of the surrounding 16 pixel region. A line of pixels at pixel number 837 counting horizontally from the left side of the clipped image, which spans the vertical extent of the image are affected by fallout from a single hot pixel. This line of pixels was removed and each pixel replaced with the mean of the pixels on either side. Each filter frame was converted from DN to radiance ($R$) using the exposure time ($t$, seconds) and updated coefficients ($C_\nu$), detailed in Table 4 of \citet{Bell2017}:

\begin{equation}
R=\frac{DN}{t}C_\nu
\end{equation}

\noindent where $C_{\nu, red}=3.56\times 10^{-7}$ W\,m$^{-2}$\,Sr$^{-1}$\,nm$^{-1}$\,DN$^{-1}$\,s, $C_{\nu, green}=3.39\times 10^{-7}$ W\,m$^{-2}$\,Sr$^{-1}$ \,nm$^{-1}$\,DN$^{-1}$\,s, and $C_{\nu, blue}=4.47\times 10^{-7}$ W\,m$^{-2}$\,Sr$^{-1}$\,nm$^{-1}$\,DN$^{-1}$\,s.

To create the RGB images shown in Fig. \ref{exampleobs}, each frame was smoothed with a 7 pixel-wide Gaussian filter and a smoothed-frame maximum calculated. The \emph{un-smoothed} frames were then scaled to the smoothed-frame maximum of the corresponding filter, so values equal to or higher than the smoothed-frame maximum were set to unity. The individual RGB frames from a given observation were combined into a single RGB image. However, the analysis described in the following sections was carried out on the individual filter frames rather than the RGB image so extinction as a function of wavelength could be investigated.

\section{Methods}

The LOS extinction within Gale Crater has been consistently monitored since sol 100 \citep{Moores2015, Moore2016, Moore2019}. \citet{Moores2015} showed that the total optical depth between the rover and the crater rim can be found using an image including a region of ground close to the rover, a section of the crater rim, and a section of sky just above the crater rim. It is assumed that the ground and the crater rim are made of the same material, that there is zero optical depth between the rover and the ground, and that the optical depth of the sky just above the crater rim is effectively infinite. This was shown to be accurate to within 4\% when using images taken between 10:00 and 14:00 Local True Solar Time (LTST, \citealp{Moores2015}). However, during the 2018/MY34 GDS, optical depths $>8$ were reported by MSL and $>10$ by Opportunity. This caused the \citet{Moores2015} method to break down: optical depths were significantly  higher than the method was valid for, foreground regions of LOS images were contaminated with substantial amounts of dust, and the crater rim was entirely obscured. Thus, an alternative method was required. 

On Earth, daytime visibility measurements (maximum distance that a feature, assumed to be black and against the horizon, is resolvable) can be used to estimate LOS extinction using Koschmieder's Law \citep{Koschmieder1924}:

\begin{equation}
\beta \,[km^{-1}] = \frac{-\ln (C_T)}{D \,[km]}
\end{equation}

\noindent where $\beta$ is the extinction coefficient (km$^{-1}$), $D$ is the distance to the furthest discernible feature (km), and $C_T$ is the contrast threshold. For the average human eye under typical lab conditions, $C_T\approx0.02$. For the Navcam images, $N/S$ (noise over signal) was 0.01-0.005, thus the observer's threshold was the limiting factor and 0.02 was used. For the Mastcam images, $N/S>0.04$ and varied significantly between filters, thus noise was the limiting factor, so $N/S$ of each filter frame was used. The impact of increasing $C_T$ is an increase in calculated extinction for a given measured distance: increasing $C_T$ from 0.02 to 0.05 increases the calculated extinction by 30\%. Conversely, decreasing $C_T$ by 50\% decreases the calculated extinction by 15\%. 

The distance to the furthest discernible feature was measured by geo-referencing with a digital terrain model (DTM) of Gale Crater using instrument pointings and rover telemetry data. The 50m/px resolution DTM of Gale Crater used \citep{Gwinner2010} was constructed from images taken by High/Super Resolution Stereo Camera (HRSC) on-board Mars Express. The DTM was limited to those regions visible to MSL when the image in question was taken, and binned and interpolated to a uniform angular resolution.  

Each image or filter, was divided into five approximately equal regions in azimuth and the furthest visible feature in each region selected. The furthest discernible feature was determined by eye, though more automated techniques were explored but ultimately found to be unfeasible. A perceptually uniform colormap was used to reduce selection biases stemming from the eye's differing sensitivity to different colors. The screen, brightness, and local illumination levels were kept as constant as possible. All images were processed three times.

Navcam LOS images suffered from a strong variation in brightness across the frame, which meant that features could not be consistently identified, leading to results which would not be representative of the actual visibility. To mitigate this, all images were averaged in azimuth to produce an average brightness curve as a function of elevation and this vertical profile was subtracted from each image. The Mastcam CRE observations, due to the noise associated with the camera, also required color stretching after the brightness correction was applied. 

Occasionally during the GDS, dust particles were deposited or removed from Navcam. These particles could mask the furthest features in an image section, rendering the visibility estimate lower than it would otherwise be, though the effect on the overall results is mitigated by using the mean of five different regions for distance determination.

MSL did not remain stationary during the GDS, causing variation in the observed scenes including changing foreground features and image rotation. For some positions, certain directions were obscured by foreground features, particularly in the south. In other cases, substantial rotation of the image resulted in some sections with no detectable features. In both these cases, distances for the affected regions were recorded as NaN and were excluded from all further analyses. 

\section{Results and Discussion}\label{discsection}

\begin{figure}
\includegraphics[width=\textwidth, trim=3cm 0cm 3cm 0cm, clip=True]{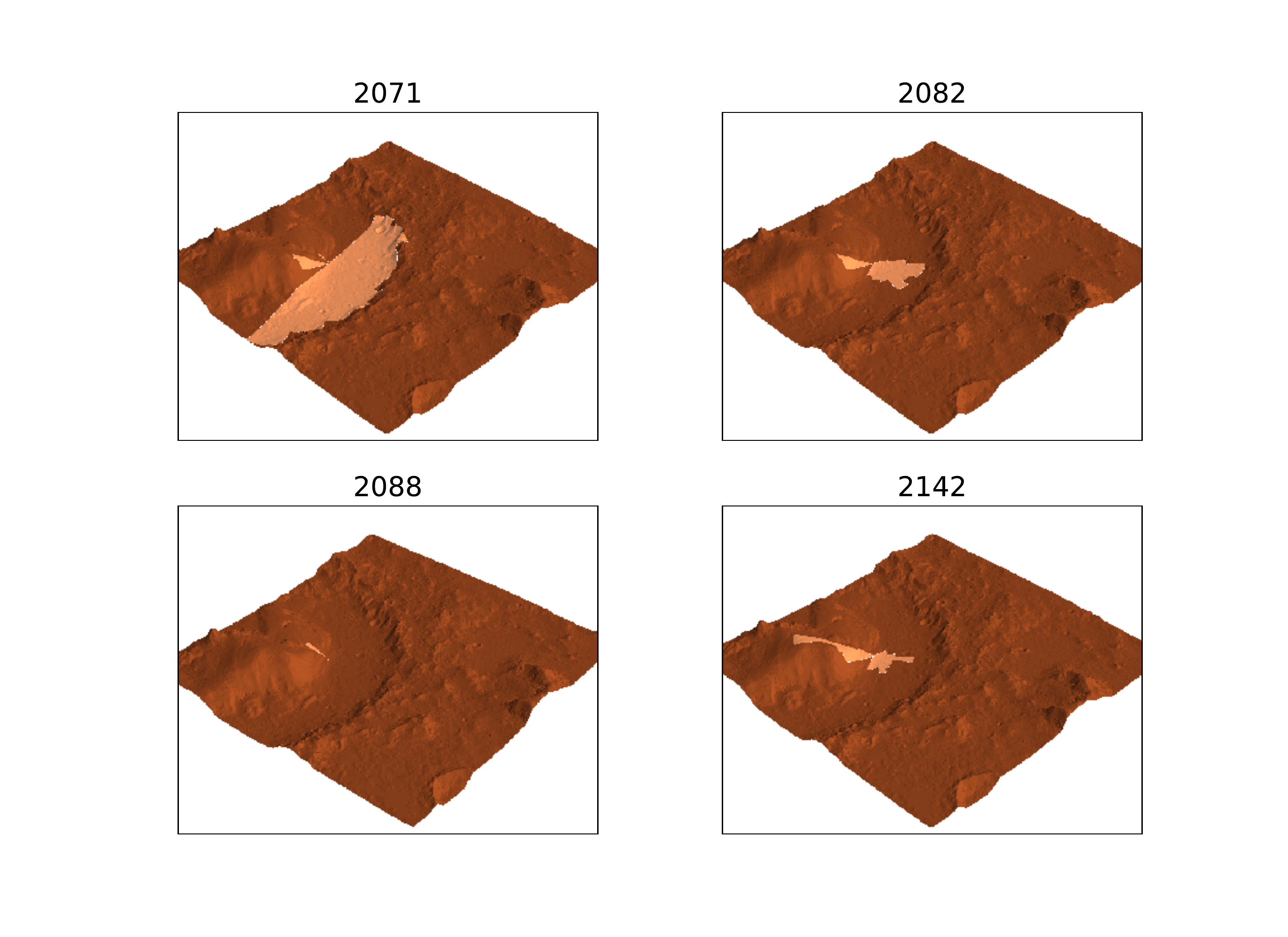}
\caption{Maximum visible distance on four sols (indicated above each panel) in all directions from DD360 observations. The lighter region overlaid on the darker Gale Crater DTM represents the area visible to MSL on that sol. North is in the lower right corner of each image and east is in the lower left corner. The northern crater rim (shown as visible in sol 2071) is approximately 30 km from MSL. On sol 2088, the maximum distance visible was less than 3 km, and thus the visible region is barely identifiable on the figure.}\label{DTM_distplot} 
\end{figure}

Figure \ref{DTM_distplot} shows the maximum distance visible in all possible directions (averaged in $15^\circ$ azimuthal bins) on four sols during the GDS from the DD360s. Dark regions of the terrain model show portions of Gale Crater that were not visible in the images from MSL, and lighter areas show those regions that were visible. 

The upper left panel shows the visibility on sol 2071 just prior to the GDS campaign initiation.  The upper right panel shows the visibility on sol 2082 during the onset of the storm. The lower left panel shows the visibility on sol 2088, around the peak of the storm. The lower right panel shows the visibility on sol 2142, during the decay phase of the storm. As can be seen, visibility reduced dramatically during the storm, with even \emph{Aeolis Mons} completely disappearing from view. The decay in extinction was substantially slower than the growth during the onset of the storm.

\begin{figure}
\includegraphics[width=\textwidth, trim=1cm 0cm 1cm 0cm, clip=True]{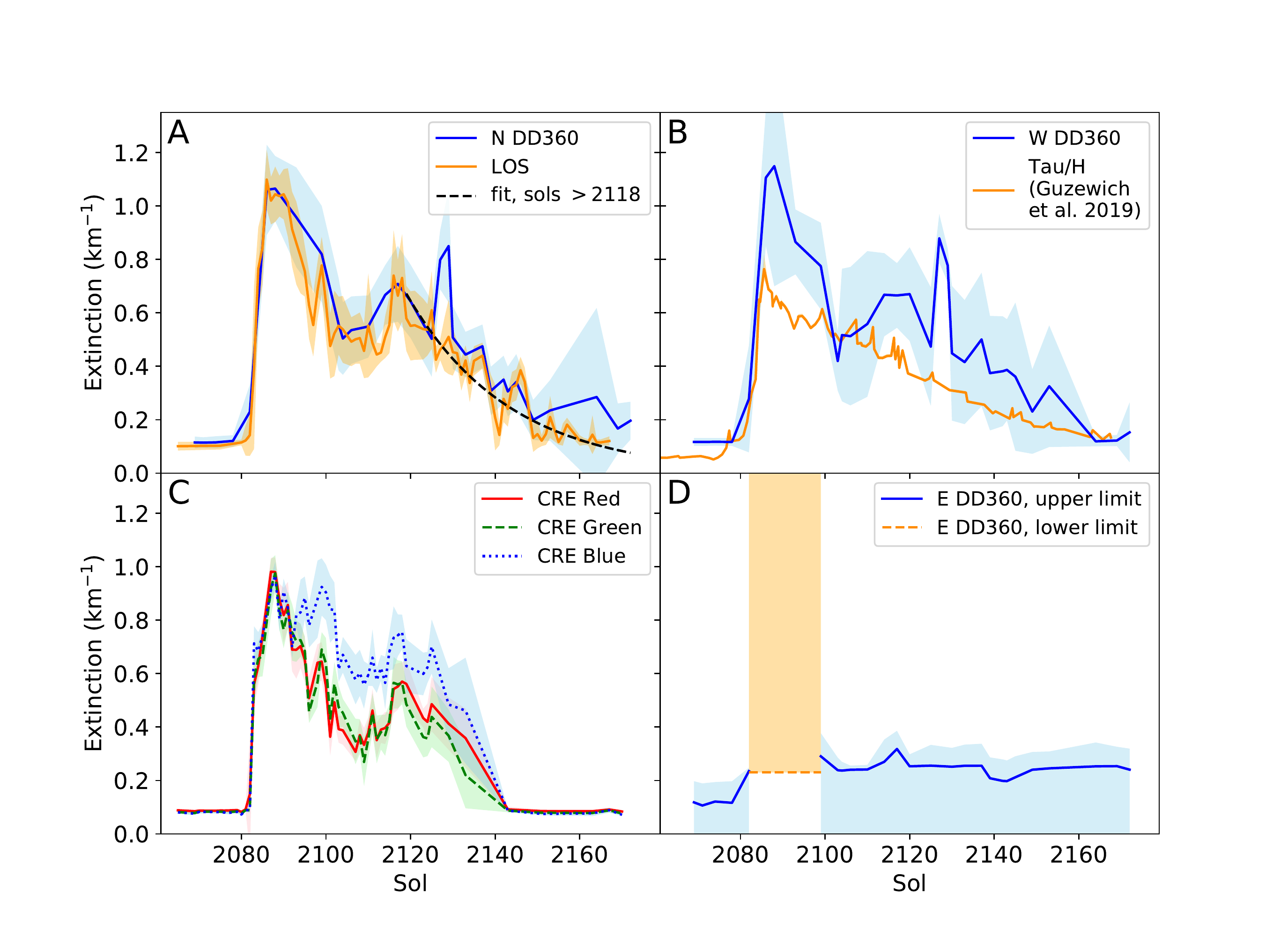}
\caption{Extinction as a function of sol. Where multiple observations were made on a given sol, the results were averaged. Panel A shows the extinction calculated from the Navcam LOS images. Panel C shows the results calculated from the Mastcam CRE images, broken down by filter as shown in the legend. Panel B shows the western DD360 extinction and the \citet{Guzewich2018} tau results for comparison, shown as tau/scale height (11.1 km). Panel D shows the limits on the eastern DD360 extinction. }\label{extvssol_multi} 
\end{figure}

Figure \ref{extvssol_multi} shows the estimated extinction for each of the image types. Panel A shows the Navcam LOS and the DD360 in the northern direction (azimuths = -45 to +45$^\circ$). Panel B shows the results from the western DD360 images (azimuths = -135 to -45$^\circ$). Panel C shows the results of the Mastcam CRE results, separated by color filter. Panel D shows the limits placed in the eastern directions from the DD360 images (azimuths = +45 to 135$^\circ$). No useful limits could be placed upon the southern directions due to the presence of foreground topography. All panels show the results averaged by sol, with the shaded regions showing the rms uncertainties from the repeated measurements. For example, for Panel A, if one Navcam LOS observation was taken on a given sol, there are 15 distances, therefore extinctions, measured (five per image, repeated three times) and the value shown is the average of all 15. The shaded uncertainty is the rms of the 15 measurements. For DD360s, the observations were divided into 90$^\circ$ azimuth bins and treated similarly. 

In northern directions, there is excellent agreement between the Navcam LOS and DD360 results. Both show a peak of $\sim$ 1.07 km$^{-1}$ between sols 2086 and 2090, with an initial decay until approximately sol 2105 (minima at extinction values of approximately 0.5 km$^{-1}$), following which an increase in extinction occurs until sol 2117, after which an approximately exponential decay occurs until the end of the GDS campaign. 

Assuming an exponential decay of the form $A\exp(-t/t_c)$, the extinction decay timescale, $t_c$, can be determined through fitting the Navcam LOS data, though the value is heavily dependent upon which time periods are fitted. For example, fitting the entire decay phase after sol 2085, including the dip through sols 2092-2118, yields a decay timescale of 90 sols. Removing the sol 2092-2118 range, as if this dip is solely an artificial effect of local dynamics and not characteristic of general dust settling and thus fitting only the initial peak and then sols $>2118$, results in a decay timescale of 43 sols which is identical to the value determined from Mastcam opacity measurements with MSL during the 2018/MY34 GDS \citep{Guzewich2018} and from the MER rovers during the 2007/MY28 GDS \citep{Lemmon2015}. Fitting solely the final decay phase after sol 2118, as shown by the black dashed line in Fig. \ref{extvssol_multi}A, yields a decay timescale of 28 sols. Due to the uncertainties associated with the measurements, the decay can also be well-represented by a linear fit. For example, fitting only the initial peak and then sols $>2118$ gives a linear fit with gradient -0.0125 km$^{-1}$sol$^{-1}$.

Mastcam red and green filters follow a very similar pattern to the Navcam observations, with a peak extinction around sol 2088, initial decay until sol 2107 (minima of approximately 0.37 km$^{-1}$), a secondary peak on sol 2117 and subsequently decaying until the end of the GDS campaign. The exact positions of the peaks are expected to vary between the three different types of observations due to differing cadences. The Mastcam CREs were taken with a substantially decreased frequency after sol 2117, thus the shape of the decay curve is difficult to constrain. The primary peak extinction value is $\sim 1.0$ km$^{-1}$ in all filters which is within uncertainties of the peak value measured by Navcam LOS and DD360 observations. However, the Mastcam blue filter shows generally higher extinction than the red or green filters during the GDS, but also had a significantly lower SNR, making it difficult to identify features, particularly in the mid-ground regions of the images. The uncertainties shown in Fig. \ref{extvssol_multi}C are the rms uncertainties from the repeated measurements, however these are likely underestimating the uncertainty for the blue filter images so should be taken with caution.
 
Western directions follow the Navcam LOS and northern DD360 results closely, with an initial peak in extinction (1.15 km$^{-1}$) on sol 2088 then decaying down to approximately 0.4 km$^{-1}$ by sol 2103, a secondary peak in extinction around sol 2117 and then a final decay phase down to normal seasonal levels. In both the northern and western DD360 observations, a third sudden peak is seen on sol 2128, though this is likely caused by foreground topography obscuring the true maximum visible distance rather than a sudden increase then decrease in extinction.

In eastern directions, with \emph{Aeolis Mons} within the FOV, only limits can be placed on the extinction. The peak of the mountain is only $\sim13$ km from the rover, thus during less dusty times, the maximum distance could not be accurately measured and the extinction is an upper limit. The base of the mountain was obscured in all images by foreground features so during dustier periods when the mountain peak was not visible, the maximum distance was not measurable. The effect of the presence of \emph{Aeolis Mons} on the maximum visible distance is clearly seen in Fig. \ref{DTM_distplot}. Interestingly, \emph{Aeolis Mons} became and remained visible earlier in the GDS decay phase than would be expected from the northern and western results, potentially due to a greater contrast between \emph{Aeolis Mons} and the background sky than that of fore-/mid-ground features against background terrain, making \emph{Aeolis Mons} easier to detect. However, it could also indicate that there is a lower dust concentration in the eastern direction or there is a gradient in dust concentration with altitude, making \emph{Aeolis Mons} easier to detect if dust is concentrated at lower altitudes.

\begin{sidewaysfigure}
\includegraphics[width=\textwidth,  trim=3cm 0cm 3cm 0cm, clip=True]{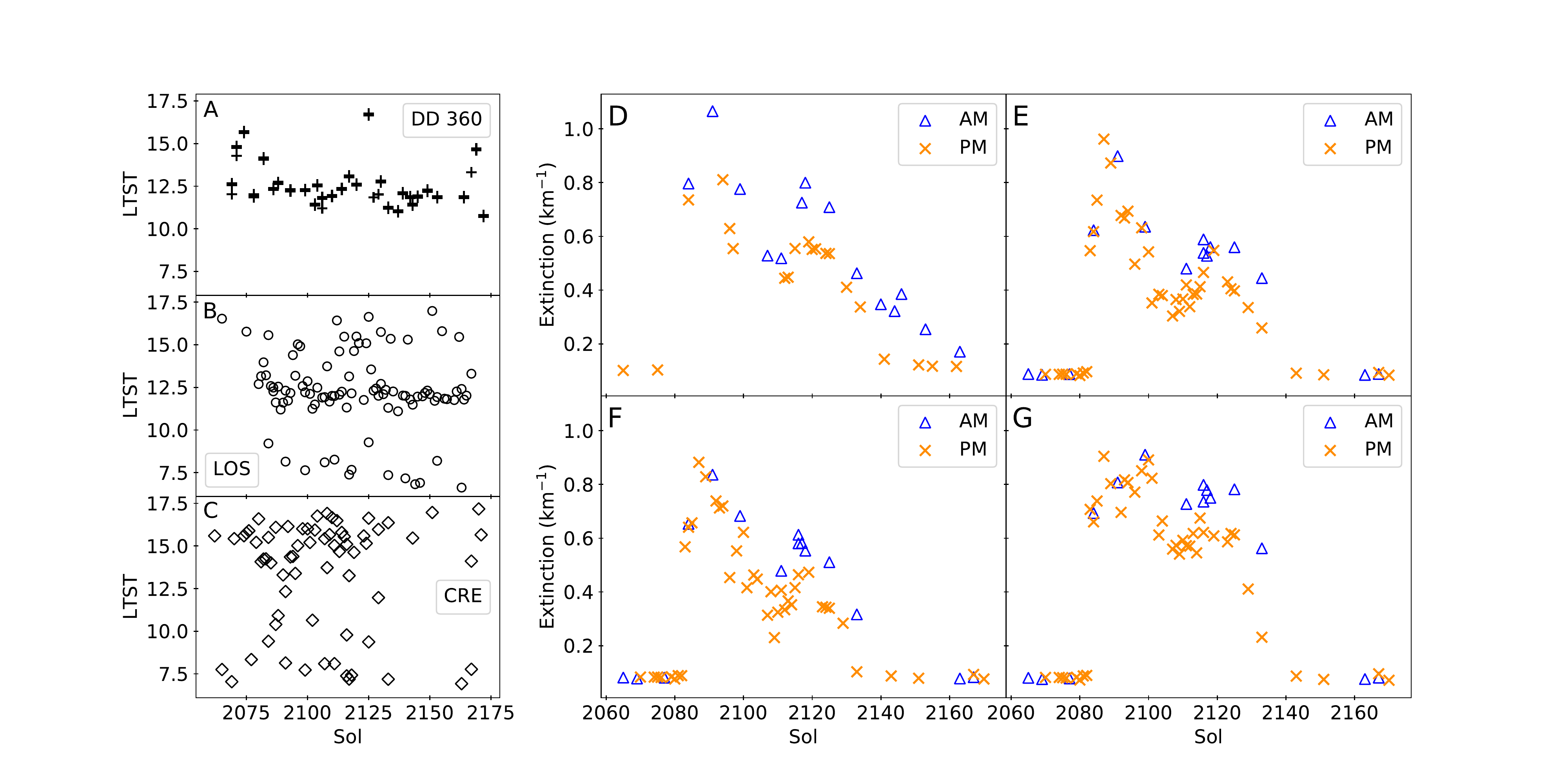}
\caption{Panels A-C show the temporal distribution of Navcam DD360, LOS, and Mastcam CRE observations. Panels D, E, F, and G show the morning (triangles) and afternoon (crosses) extinction results from the  Navcam LOS, Mastcam CRE red, green, and blue observations respectively.}\label{obsdistribution}
\end{sidewaysfigure}

As shown in the left hand panels of Fig. \ref{obsdistribution}, the observations were taken over a variety of times during the GDS. Navcam LOS are primarily clustered around local noon $\pm 2$ hours and a smaller number taken in the morning and afternoon. Navcam DD360s were almost exclusively taken around noon, with a few observations obtained in the afternoon. Mastcam CREs were most often obtained in the afternoon with a sizable number taken in the morning and very few taken around noon. LOS and CRE observations were obtained twice or more on 15 and 11 sols respectively. Panels D-G of Fig. \ref{obsdistribution} show the results of the LOS and CRE observations taken only in the morning (prior to 10:00 LTST) or afternoon (after 14:00 LTST), averaged by time-frame and by sol. The morning extinction estimates are consistently higher than those in the afternoon. Comparing with the mid-sol time-frame (noon $\pm 2$ hours), afternoon extinction estimates are generally lower and morning results generally higher. This trend is also seen on sols with multiple observations on a single sol: almost exclusively the morning results are higher than the mid-sol or afternoon results. If this were a lighting effect, given the northern pointing of the CRE and LOS observations, the afternoon and morning results would be the same, however that is not observed. This may be indicative of mixing within Gale Crater; overnight, conditions are stably stratified and dust should primarily settle, resulting in a thinner layer of higher opacity at sunrise. During the day, mixing redistributes this dust through a thicker layer (4-6 km in height) of lower opacity. However, this effect is also observed in column opacity measurements \citep[e.g.][]{Guzewich2018} which would not be affected substantially by atmospheric mixing thus additional work is required to reliably determine the cause of diurnal variation.

Comparing the line-of-sight extinction estimates to the column opacity measurements from Mastcam sun imaging \citep[shown as tau/H on Fig. \ref{extvssol_multi}B of this work]{Guzewich2018} show some agreement and some differences. The column opacity increases by an order of magnitude during the initial onset of the storm, which is reflected in the LOS measurements (an increase from $<0.1$ to $>1.0$ km$^{-1}$). The position of the peak is earlier in the column opacity measurements than in the Navcam and Mastcam measurements, though not substantially so: the column opacity measurements place the peak at approximately 2085, whereas the Mastcam CRE and Navcam LOS and DD360 results place it at approximately 2088 and in the range 2086-2090 respectively.  \citet{Guzewich2018} also report that morning opacities are higher than those taken in the afternoon or mid-sol time-frames, which is in agreement with the results of our work. However, despite our results showing a substantial dip in extinction around sol 2105, no such dip is seen in the column opacities from \citet{Guzewich2018}. This could represent two waves of dust coming into the crater. Alternatively, a temporary increase in dust-lifting that was constrained to the very near-surface, or a change in dynamics within Gale Crater compacting the dust layer, could explain this difference. In any of these cases, the overall effect on the vertical column opacity would be negligible. 

\section{Conclusions}

LOS extinction estimates have been determined using Navcam and Mastcam observations taken throughout the 2018/MY34 GDS. The peak extinction is an order of magnitude higher than previous extinction measurements made at Gale Crater, in agreement with column opacity measurements \citep{Guzewich2018}. Northern and western results agree well; east-pointing results show lower extinction than expected, indicating either directional variation in extinction or a easier detectability of \emph{Aeolis Mons} against the background sky. Mastcam red and green filter results agree, but blue filter results show higher extinction, likely an effect of higher noise levels in that filter. Diurnally, morning results show higher extinction than afternoon results, potentially indicative of atmospheric mixing though additional work is required before a definitive cause can be identified.

\acknowledgments

Smith, Moores, and Moore were supported by the Canadian Space Agency's MSL participating scientist program. Guzewich was supported by the MSL Participating Scientist program. MSL data are provided via the PDS 6 months after receipt on Earth. Visibility and extinction measurements are archived via Mendeley (doi:10.17632/pxnddn22gn.1). Smith would like to acknowledge J. Van Beek for helpful discussions of Mastcam data.  
\bibliography{los.bib}

\begin{thebibliography}{}

\bibitem [\protect \citeauthoryear {%
Bell%
\ \protect \BOthers {.}}{%
Bell%
\ \protect \BOthers {.}}{%
{\protect \APACyear {2017}}%
}]{%
Bell2017}
\APACinsertmetastar {%
Bell2017}%
\begin{APACrefauthors}%
Bell, J\BPBI F.%
, Godber, A.%
, McNair, S.%
, Caplinger, M\BPBI A.%
, Maki, J\BPBI N.%
, Lemmon, M\BPBI T.%
\BDBL {}Deen, R\BPBI G.%
\end{APACrefauthors}%
\unskip\
\newblock
\APACrefYearMonthDay{2017}{7}{}.
\newblock
{\BBOQ}\APACrefatitle {The Mars Science Laboratory Curiosity rover Mastcam
  instruments: Preflight and in‐flight calibration, validation, and data
  archiving} {The mars science laboratory curiosity rover mastcam instruments:
  Preflight and in‐flight calibration, validation, and data
  archiving}.{\BBCQ}
\newblock
\APACjournalVolNumPages{Earth and Space Science}{4}{7}{396--452}.
\newblock
\begin{APACrefURL} \url{https://doi.org/10.1002/2016EA000219} \end{APACrefURL}
\newblock
\begin{APACrefDOI} \doi{10.1002/2016EA000219} \end{APACrefDOI}
\PrintBackRefs{\CurrentBib}

\bibitem [\protect \citeauthoryear {%
Clancy%
\ \protect \BOthers {.}}{%
Clancy%
\ \protect \BOthers {.}}{%
{\protect \APACyear {2010}}%
}]{%
Clancy2010}
\APACinsertmetastar {%
Clancy2010}%
\begin{APACrefauthors}%
Clancy, R\BPBI T.%
, Wolff, M\BPBI J.%
, Whitney, B\BPBI A.%
, Cantor, B\BPBI A.%
, Smith, M\BPBI D.%
\BCBL {}\ \BBA {} McConnochie, T\BPBI H.%
\end{APACrefauthors}%
\unskip\
\newblock
\APACrefYearMonthDay{2010}{}{}.
\newblock
{\BBOQ}\APACrefatitle {Extension of atmospheric dust loading to high altitudes
  during the 2001 Mars dust storm: MGS TES limb observations} {Extension of
  atmospheric dust loading to high altitudes during the 2001 mars dust storm:
  Mgs tes limb observations}.{\BBCQ}
\newblock
\APACjournalVolNumPages{Icarus}{207}{1}{98 - 109}.
\newblock
\begin{APACrefURL}
  \url{http://www.sciencedirect.com/science/article/pii/S0019103509004278}
  \end{APACrefURL}
\newblock
\begin{APACrefDOI} \doi{https://doi.org/10.1016/j.icarus.2009.10.011}
  \end{APACrefDOI}
\PrintBackRefs{\CurrentBib}

\bibitem [\protect \citeauthoryear {%
Guzewich%
\ \protect \BOthers {.}}{%
Guzewich%
\ \protect \BOthers {.}}{%
{\protect \APACyear {2019}}%
}]{%
Guzewich2018}
\APACinsertmetastar {%
Guzewich2018}%
\begin{APACrefauthors}%
Guzewich, S\BPBI D.%
, Lemmon, M.%
, Smith, C\BPBI L.%
, Martínez, G.%
, de Vicente-Retortillo, Ã.%
, Newman, C\BPBI E.%
\BDBL {}Zorzano~Mier, M\BHBI P.%
\end{APACrefauthors}%
\unskip\
\newblock
\APACrefYearMonthDay{2019}{}{}.
\newblock
{\BBOQ}\APACrefatitle {Mars Science Laboratory Observations of the 2018/Mars
  Year 34 Global Dust Storm} {Mars science laboratory observations of the
  2018/mars year 34 global dust storm}.{\BBCQ}
\newblock
\APACjournalVolNumPages{Geophysical Research Letters}{46}{1}{71-79}.
\newblock
\begin{APACrefURL}
  \url{https://agupubs.onlinelibrary.wiley.com/doi/abs/10.1029/2018GL080839}
  \end{APACrefURL}
\newblock
\begin{APACrefDOI} \doi{10.1029/2018GL080839} \end{APACrefDOI}
\PrintBackRefs{\CurrentBib}

\bibitem [\protect \citeauthoryear {%
{Gwinner}%
\ \protect \BOthers {.}}{%
{Gwinner}%
\ \protect \BOthers {.}}{%
{\protect \APACyear {2010}}%
}]{%
Gwinner2010}
\APACinsertmetastar {%
Gwinner2010}%
\begin{APACrefauthors}%
{Gwinner}, K.%
, {Scholten}, F.%
, {Preusker}, F.%
, {Elgner}, S.%
, {Roatsch}, T.%
, {Spiegel}, M.%
\BDBL {}{Heipke}, C.%
\end{APACrefauthors}%
\unskip\
\newblock
\APACrefYearMonthDay{2010}{}{}.
\newblock
{\BBOQ}\APACrefatitle {{Topography of Mars from global mapping by HRSC
  high-resolution digital terrain models and orthoimages: Characteristics and
  performance}} {{Topography of Mars from global mapping by HRSC
  high-resolution digital terrain models and orthoimages: Characteristics and
  performance}}.{\BBCQ}
\newblock
\APACjournalVolNumPages{Earth and Planetary Science Letters}{294}{}{506-519}.
\newblock
\begin{APACrefDOI} \doi{10.1016/j.epsl.2009.11.007} \end{APACrefDOI}
\PrintBackRefs{\CurrentBib}

\bibitem [\protect \citeauthoryear {%
{Koschmieder}%
}{%
{Koschmieder}%
}{%
{\protect \APACyear {1924}}%
}]{%
Koschmieder1924}
\APACinsertmetastar {%
Koschmieder1924}%
\begin{APACrefauthors}%
{Koschmieder}, H.%
\end{APACrefauthors}%
\unskip\
\newblock
\APACrefYearMonthDay{1924}{}{}.
\newblock
{\BBOQ}\APACrefatitle {{Theorie der horizontalen Sichweite}} {{Theorie der
  horizontalen Sichweite}}.{\BBCQ}
\newblock
\APACjournalVolNumPages{Beitr. z. Phys. d. freien Atm.}{12}{}{171-181}.
\PrintBackRefs{\CurrentBib}

\bibitem [\protect \citeauthoryear {%
{Lemmon}%
\ \protect \BOthers {.}}{%
{Lemmon}%
\ \protect \BOthers {.}}{%
{\protect \APACyear {2017}}%
}]{%
Lemmon2017}
\APACinsertmetastar {%
Lemmon2017}%
\begin{APACrefauthors}%
{Lemmon}, M\BPBI T.%
, {Newman}, C\BPBI E.%
, {Renno}, N.%
, {Mason}, E.%
, {Battalio}, M.%
, {Richardson}, M\BPBI I.%
\BCBL {}\ \BBA {} {Kahanp{\"a}{\"a}}, H.%
\end{APACrefauthors}%
\unskip\
\newblock
\APACrefYearMonthDay{2017}{{\APACmonth{03}}}{}.
\newblock
{\BBOQ}\APACrefatitle {{Dust Devil Activity at the Curiosity Mars Rover Field
  Site}} {{Dust Devil Activity at the Curiosity Mars Rover Field Site}}.{\BBCQ}
\newblock
\BIn{} \APACrefbtitle {Lunar and Planetary Science Conference} {Lunar and
  planetary science conference}\ (\BVOL~48, \BPG~2952).
\PrintBackRefs{\CurrentBib}

\bibitem [\protect \citeauthoryear {%
Lemmon%
\ \protect \BOthers {.}}{%
Lemmon%
\ \protect \BOthers {.}}{%
{\protect \APACyear {2015}}%
}]{%
Lemmon2015}
\APACinsertmetastar {%
Lemmon2015}%
\begin{APACrefauthors}%
Lemmon, M\BPBI T.%
, Wolff, M\BPBI J.%
, Bell, J\BPBI F.%
, Smith, M\BPBI D.%
, Cantor, B\BPBI A.%
\BCBL {}\ \BBA {} Smith, P\BPBI H.%
\end{APACrefauthors}%
\unskip\
\newblock
\APACrefYearMonthDay{2015}{}{}.
\newblock
{\BBOQ}\APACrefatitle {Dust aerosol, clouds, and the atmospheric optical depth
  record over 5 Mars years of the Mars Exploration Rover mission} {Dust
  aerosol, clouds, and the atmospheric optical depth record over 5 mars years
  of the mars exploration rover mission}.{\BBCQ}
\newblock
\APACjournalVolNumPages{Icarus}{251}{}{96 - 111}.
\newblock
\begin{APACrefURL}
  \url{http://www.sciencedirect.com/science/article/pii/S0019103514001559}
  \end{APACrefURL}
\newblock
\APACrefnote{Dynamic Mars}
\newblock
\begin{APACrefDOI} \doi{https://doi.org/10.1016/j.icarus.2014.03.029}
  \end{APACrefDOI}
\PrintBackRefs{\CurrentBib}

\bibitem [\protect \citeauthoryear {%
Malvar%
, He%
\BCBL {}\ \BBA {} Cutler%
}{%
Malvar%
\ \protect \BOthers {.}}{%
{\protect \APACyear {2004}}%
}]{%
Malvar2004}
\APACinsertmetastar {%
Malvar2004}%
\begin{APACrefauthors}%
Malvar, H.%
, He, L\BHBI w.%
\BCBL {}\ \BBA {} Cutler, R.%
\end{APACrefauthors}%
\unskip\
\newblock
\APACrefYearMonthDay{2004}{06}{}.
\newblock
{\BBOQ}\APACrefatitle {High-quality linear interpolation for demosaicing of
  Bayer-patterned color images} {High-quality linear interpolation for
  demosaicing of bayer-patterned color images}.{\BBCQ}
\newblock
\BIn{} (\BVOL~3, \BPG~iii - 485).
\newblock
\begin{APACrefDOI} \doi{10.1109/ICASSP.2004.1326587} \end{APACrefDOI}
\PrintBackRefs{\CurrentBib}

\bibitem [\protect \citeauthoryear {%
{Moore}%
\ \protect \BOthers {.}}{%
{Moore}%
\ \protect \BOthers {.}}{%
{\protect \APACyear {2016}}%
}]{%
Moore2016}
\APACinsertmetastar {%
Moore2016}%
\begin{APACrefauthors}%
{Moore}, C\BPBI A.%
, {Moores}, J\BPBI E.%
, {Lemmon}, M\BPBI T.%
, {Rafkin}, S\BPBI C\BPBI R.%
, {Francis}, R.%
, {Pla-Garc{\'{\i}}a}, J.%
\BDBL {}{Burton}, J\BPBI R.%
\end{APACrefauthors}%
\unskip\
\newblock
\APACrefYearMonthDay{2016}{}{}.
\newblock
{\BBOQ}\APACrefatitle {{A full martian year of line-of-sight extinction within
  Gale Crater, Mars as acquired by the MSL Navcam through sol 900}} {{A full
  martian year of line-of-sight extinction within Gale Crater, Mars as acquired
  by the MSL Navcam through sol 900}}.{\BBCQ}
\newblock
\APACjournalVolNumPages{Icarus}{264}{}{102-108}.
\PrintBackRefs{\CurrentBib}

\bibitem [\protect \citeauthoryear {%
Moore%
\ \protect \BOthers {.}}{%
Moore%
\ \protect \BOthers {.}}{%
{\protect \APACyear {2019}}%
}]{%
Moore2019}
\APACinsertmetastar {%
Moore2019}%
\begin{APACrefauthors}%
Moore, C\BPBI A.%
, Moores, J\BPBI E.%
, Newman, C\BPBI E.%
, Lemmon, M\BPBI T.%
, Guzewich, S\BPBI D.%
\BCBL {}\ \BBA {} Battalio, M.%
\end{APACrefauthors}%
\unskip\
\newblock
\APACrefYearMonthDay{2019}{}{}.
\newblock
{\BBOQ}\APACrefatitle {Vertical and horizontal heterogeneity of atmospheric
  dust loading in northern Gale Crater, Mars} {Vertical and horizontal
  heterogeneity of atmospheric dust loading in northern gale crater,
  mars}.{\BBCQ}
\newblock
\APACjournalVolNumPages{Icarus}{329}{}{197 - 206}.
\newblock
\begin{APACrefURL}
  \url{http://www.sciencedirect.com/science/article/pii/S0019103517306000}
  \end{APACrefURL}
\newblock
\begin{APACrefDOI} \doi{https://doi.org/10.1016/j.icarus.2019.03.041}
  \end{APACrefDOI}
\PrintBackRefs{\CurrentBib}

\bibitem [\protect \citeauthoryear {%
{Moores}%
\ \protect \BOthers {.}}{%
{Moores}%
\ \protect \BOthers {.}}{%
{\protect \APACyear {2015}}%
}]{%
Moores2015}
\APACinsertmetastar {%
Moores2015}%
\begin{APACrefauthors}%
{Moores}, J\BPBI E.%
, {Lemmon}, M\BPBI T.%
, {Kahanp{\"a}{\"a}}, H.%
, {Rafkin}, S\BPBI C\BPBI R.%
, {Francis}, R.%
, {Pla-Garcia}, J.%
\BDBL {}{McCullough}, E.%
\end{APACrefauthors}%
\unskip\
\newblock
\APACrefYearMonthDay{2015}{}{}.
\newblock
{\BBOQ}\APACrefatitle {{Observational evidence of a suppressed planetary
  boundary layer in northern Gale Crater, Mars as seen by the Navcam instrument
  onboard the Mars Science Laboratory rover}} {{Observational evidence of a
  suppressed planetary boundary layer in northern Gale Crater, Mars as seen by
  the Navcam instrument onboard the Mars Science Laboratory rover}}.{\BBCQ}
\newblock
\APACjournalVolNumPages{Icarus}{249}{}{129-142}.
\newblock
\begin{APACrefDOI} \doi{10.1016/j.icarus.2014.09.020} \end{APACrefDOI}
\PrintBackRefs{\CurrentBib}

\bibitem [\protect \citeauthoryear {%
Smith%
, Zorzano%
, Lemmon%
, Martín-Torres%
\BCBL {}\ \BBA {} de Cal%
}{%
Smith%
\ \protect \BOthers {.}}{%
{\protect \APACyear {2016}}%
}]{%
Smith2016}
\APACinsertmetastar {%
Smith2016}%
\begin{APACrefauthors}%
Smith, M\BPBI D.%
, Zorzano, M\BHBI P.%
, Lemmon, M.%
, Martín-Torres, J.%
\BCBL {}\ \BBA {} de Cal, T\BPBI M.%
\end{APACrefauthors}%
\unskip\
\newblock
\APACrefYearMonthDay{2016}{}{}.
\newblock
{\BBOQ}\APACrefatitle {Aerosol optical depth as observed by the Mars Science
  Laboratory REMS UV photodiodes} {Aerosol optical depth as observed by the
  mars science laboratory rems uv photodiodes}.{\BBCQ}
\newblock
\APACjournalVolNumPages{Icarus}{280}{}{234 - 248}.
\newblock
\begin{APACrefURL}
  \url{http://www.sciencedirect.com/science/article/pii/S0019103516304158}
  \end{APACrefURL}
\newblock
\APACrefnote{MicroMars to MegaMars}
\newblock
\begin{APACrefDOI} \doi{https://doi.org/10.1016/j.icarus.2016.07.012}
  \end{APACrefDOI}
\PrintBackRefs{\CurrentBib}

\bibitem [\protect \citeauthoryear {%
Viúdez-Moreiras%
\ \protect \BOthers {.}}{%
Viúdez-Moreiras%
\ \protect \BOthers {.}}{%
{\protect \APACyear {{\protect \BIP {}}}}%
}]{%
Viudez2018}
\APACinsertmetastar {%
Viudez2018}%
\begin{APACrefauthors}%
Viúdez-Moreiras, D.%
, Newman, C.%
, de~la Torre, M.%
, Martínez, G.%
, Guzewich, S.%
, Lemmon, M.%
\BDBL {}Gómez-Elvira, J.%
\end{APACrefauthors}%
\unskip\
\newblock
\APACrefYearMonthDay{{\protect \BIP {}}}{}{}.
\newblock
{\BBOQ}\APACrefatitle {Effects of the MY34/2018 Global Dust Storm as Measured
  by MSL REMS in Gale Crater} {Effects of the my34/2018 global dust storm as
  measured by msl rems in gale crater}.{\BBCQ}
\newblock
\APACjournalVolNumPages{Journal of Geophysical Research: Planets}{}{}{}.
\newblock
\begin{APACrefURL}
  \url{https://agupubs.onlinelibrary.wiley.com/doi/abs/10.1029/2019JE005985}
  \end{APACrefURL}
\newblock
\begin{APACrefDOI} \doi{10.1029/2019JE005985} \end{APACrefDOI}
\PrintBackRefs{\CurrentBib}

\bibitem [\protect \citeauthoryear {%
{Zurek}%
\ \BBA {} {Martin}%
}{%
{Zurek}%
\ \BBA {} {Martin}%
}{%
{\protect \APACyear {1993}}%
}]{%
Zurek1993}
\APACinsertmetastar {%
Zurek1993}%
\begin{APACrefauthors}%
{Zurek}, R\BPBI W.%
\BCBT {}\ \BBA {} {Martin}, L\BPBI J.%
\end{APACrefauthors}%
\unskip\
\newblock
\APACrefYearMonthDay{1993}{{\APACmonth{02}}}{}.
\newblock
{\BBOQ}\APACrefatitle {{Interannual variability of planet-encircling dust
  storms on Mars}} {{Interannual variability of planet-encircling dust storms
  on Mars}}.{\BBCQ}
\newblock
\APACjournalVolNumPages{Journal of Geophysical Research,}{98}{}{3247-3259}.
\newblock
\begin{APACrefDOI} \doi{10.1029/92JE02936} \end{APACrefDOI}
\PrintBackRefs{\CurrentBib}

\end{thebibliography}

\end{document}